\documentclass[twocolumn,english,aps,prb,showpacs,superscriptaddress,floats,amsmath,amssymb,floatfix]{revtex4}

\usepackage[latin9]{inputenc}
\usepackage{color}
\usepackage{amsmath}
\usepackage{graphicx}
\usepackage{graphics}
\usepackage{gensymb}
\usepackage{pdftexcmds}
\usepackage{ifpdf}
\usepackage{amssymb}
\usepackage{subcaption}
\usepackage{bm}
\usepackage{dsfont}

\makeatletter
\@ifundefined{textcolor}{}
{%
 \definecolor{BLACK}{gray}{0}
 \definecolor{WHITE}{gray}{1}
 \definecolor{RED}{rgb}{1,0,0}
 \definecolor{GREEN}{rgb}{0,1,0}
 \definecolor{GREEN2}{rgb}{0,0.4,0}
 \definecolor{BLUE}{rgb}{0,0,1}
 \definecolor{CYAN}{cmyk}{1,0,0,0}
 \definecolor{MAGENTA}{cmyk}{0,1,0,0}
 \definecolor{YELLOW}{cmyk}{0,0,1,0}
 \definecolor{YELLOW2}{cmyk}{0,0,1,0.6}
 \definecolor{ORANGE}{rgb}{1,0.22,0}

 }

\def \be {\begin{equation}}
\def \ee {\end{equation}}
\def \bea {\begin{eqnarray}}
\def \eea {\end{eqnarray}}

\allowdisplaybreaks
\makeatother

\begin{document}

\title{Anomalous Hall effect in the coplanar antiferromagnetic coloring-triangular lattice}

\author{A. C. Duran}
\affiliation{Instituto de F\'isica de L\'iquidos y Sistemas Biol\'ogicos, CCT La Plata, CONICET and Departamento de F\'isica, Facultad de Ciencias Exactas, Universidad Nacional de La Plata, C.C. 67, 1900 La Plata, Argentina}
\author{S.A. Osorio}
\affiliation{Instituto de Nanociencia y Nanotecnolog\'ia, CNEA-CONICET, Centro At\'omico Bariloche, (R8402AGP) S. C. de Bariloche, R\'io Negro, Argentina}
\author{M.B. Sturla}
\affiliation{Instituto de F\'isica de L\'iquidos y Sistemas Biol\'ogicos, CCT La Plata, CONICET and Departamento de F\'isica, Facultad de Ciencias Exactas, Universidad Nacional de La Plata, C.C. 67, 1900 La Plata, Argentina}
%
%
\begin{abstract}
 We study the anomalous Hall effect on the antiferromagnetic coloring-triangular lattice with a coplanar magnetic configuration in presence of a spin-orbit interaction. The effect of the spin-orbit coupling is included at effective level as a rotation of the electronic spin as the electrons hop from site to site. Our result reveals that a finite Hall conductivity in the planar 120$\degree$ structure takes place if a finite spin-orbit coupling is present. A quantized Hall conductivity occurs at global band gaps resulting from the topologically non-trivial band structure.
\end{abstract}

\maketitle

\section{Introduction}
\label{sec:intro}
%

%

The anomalous Hall effect (AHE) is characterized by the presence of a Hall conductivity in absence of an external magnetic field, in systems where the time-reversal symmetry is broken, as for example, in magnetic states with net magnetization.
But even in absence of external field and net magnetization the Hall phenomena can be observed. In non-collinear magnetic textures, like ferromagnetic and antiferromagnetic skyrmion lattices \cite{nagaosa2013topological,rosales2015three,osorio2017composite,tome2021topological}, the so-called topological Hall effect \cite{nagaosa2013topological} can be observed. Despite their distinctive features, in both cases the Hall phenomena can be described by a non-trivial Berry phase acquired by the electronic states, that leads to a non-zero Berry curvature.  
In the AHE the underlying physics can be described as a consequence of a reciprocal-space Berry curvature while in the case of the topological Hall effect as consequence of a real-space Berry curvature.

When the magnetic configuration corresponds to coplanar structure with no net magnetization it is not obvious that a Hall effect can take place. While the coplanar configuration seems to fail generating a real-space Berry curvature, the lack of net magnetization does in the reciprocal space. 

However it is well known that certain frustrated magnets with coplanar non-collinear states, such as Mn$_3$Sn \cite{nakatsuji2015large} and Mn$_3$Ge \cite{nayak2016large}, exhibit a large value for the Hall conductivity \cite{chen2014anomalous,kubler2014non,zhang2017strong}. Here, as in the case of a collinear state, the spin-orbit coupling (SOC) plays a central role in the electronic properties of these systems and in particular in the Hall conductivity.

The AHE in coplanar antiferromagnets suggest the presence of a Berry curvature associated to the SOC. Indeed, it turns out that the effect of the  SOC can be interpreted in terms of an effective magnetization configuration with a net contribution to the Berry curvature \cite{busch2020microscopic}.

Recently, Zhang et. al in Ref. \onlinecite{zhang2020real} have shown that introducing the SOC at an effective level in the hopping amplitudes, a direct contribution to the Berry phase in the reciprocal space can be obtained, leading to a non-zero Berry curvature when the SOC is adequately chosen.     
The effect of the SOC is taken into account by a proper change of the hopping terms. To this end they introduce a set of SU(2) rotation matrices $U_{i,j}$ that take into account the rotation induced by the SOC when the electrons hop between sites ``i'' and ``j''.

In this work we will analyze the AHE on a color-triangular (CT) lattice with a coplanar magnetic configuration, with no net magnetization.  
The CT lattice is a triangular lattice with non-uniform NN hoppings that can be mapped to a three-colored triangular tessellation which exhibits a band structure quite similar to that of the kagome lattice \cite{zhang2019kagome}. 
As is well know many interesting phenomena are attached to the kagome lattice, such as frustration, Dirac cones, and flat bands, which are the scenario for strongly correlated phases \cite{ohgushi2000spin,balents2010spin,zhang2005experimental,kane2005quantum,guo2009topological,zheng2014exotic,liu2013flat,tang2011high}. 
Due to the aforementioned connection, many of these phenomena are expected to be present in the CT lattices. 
Recently, it was shown that the organic compound Cu-dicyanobenzene, that presents the structure of the CT lattice and a ferromagnetic state, exhibits the AHE \cite{gao2020quantum}. In this system, a band structure with non-trivial Berry curvatures emerges as a consequence of a net magnetized state, the presence of a SOC, and the similarity with its kagome partner.  
To the best of our knowldege, the antiferromagnetic counterpart of these CT lattices was not yet studied and are the main objective of this article. In what follows we show that the AHE could emerge in the $120\degree$ coplanar structure without net magnetization, which is characteristic of the triangular antiferromagnets. In the model discussed here, proper combinations of the hopping strengths and, crucially, the SOC, lead to a significant contributions to the AHE even in the absence of a ferromagnetic order. 
  As we show here, in this model the interplay between SOC and inhomogeneous hoping integrals leads to the appearance of topologically non trivial bands. Moreover, for a suitable set of parameters a significant contribution to the Hall conductivity is present. 
  We show that, as a consequence of the presence of topologically non-trivial bands with a definite Chern number, a quantized AHE takes place within the global band-gaps.

  The article is organized as follows. In Sec. \ref{sec:model} 
  we introduce the model under study and its band structure. In Sec. \ref{sec:phase_diagram} we analyze the band topology and deliver the main results on the Hall conductivity and the presence of chiral edge states. In Sec. \ref{sec:realizations} we discuss possible experimental realizations of the model as considered here. Finally, in Sec. \ref{sec:conclusions}, we summarize our main results and conclusions.

\section{The model and band structure}
\label{sec:model}
\begin{figure}[htb]
\includegraphics[width=8.5cm]{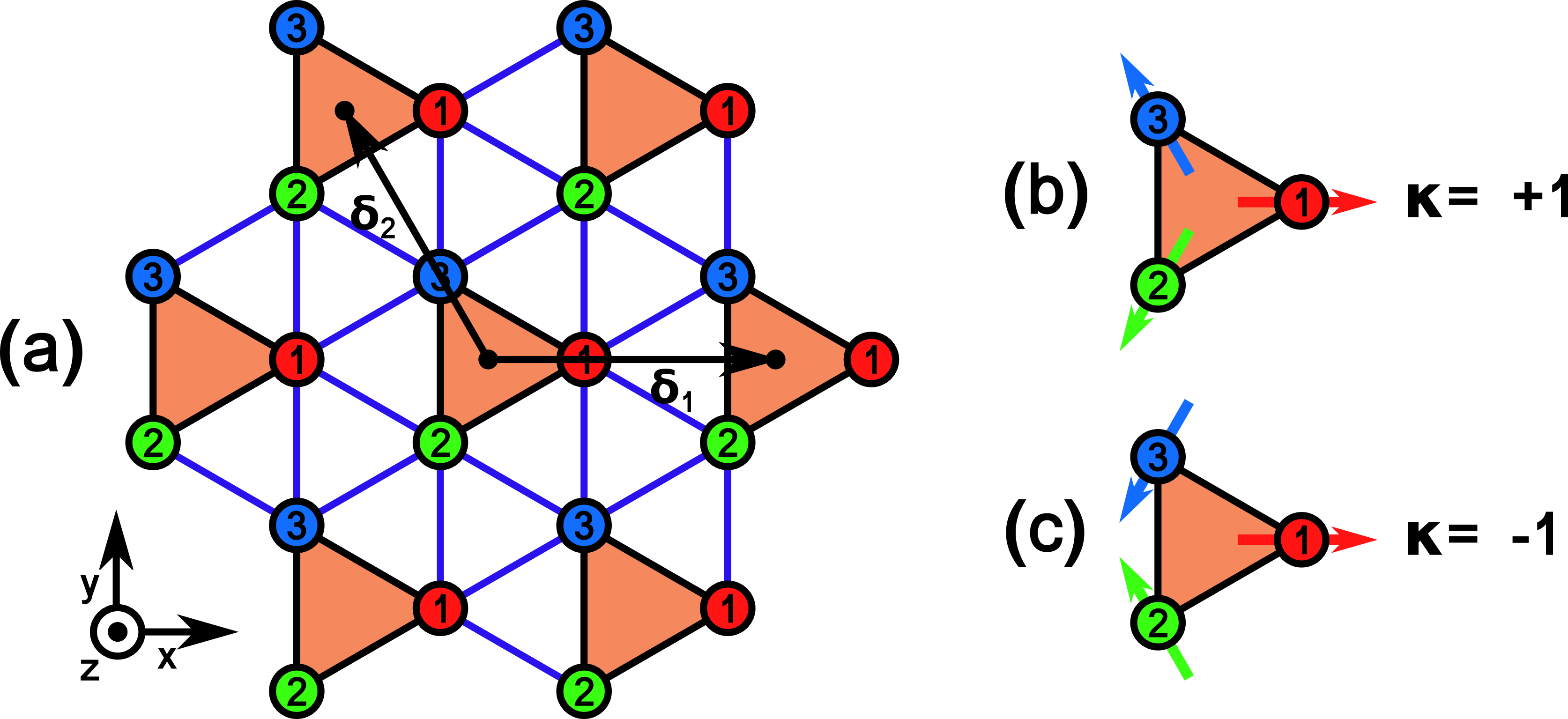}
\caption{(a) The triangular lattice with the plaquettes highlighted in orange representing the magnetic cell. The intracell links (black links) include the $t_1$ hopping and the SOC while the intercell links corresponds to the $t_2$ hopping. The spins on each site are labeled by 1, 2 and 3. The vectors $\boldsymbol{\delta}_{1,2}$ represent the primitive lattice vectors of the Bravais lattice. For the 120$\degree$ structure ((b) and (c)), two possible chiralities are compatible with this magnetic ordering.}
\label{fig:red}
\end{figure}

The model consists of a 2D triangular lattice with a lattice spacing $a$. The magnetic cell is comprised of three sites where the spins lie in the plane of the system forming a 120$\degree$ structure (Fig. \ref{fig:red}). The bonds connecting two sites within a magnetic cell (black lines in Fig. \ref{fig:red} (a)) are characterized by a hopping amplitude $t_1$ and the SOC. Whereas the bonds connecting sites between adjacent magnetic cells are characterized by a hopping amplitude $t_2$ without SOC (blue lines in Fig. \ref{fig:red} (a)). The effect of the magnetic texture is taken into account via a Hund coupling between the electronic spin and the localized magnetic moments. The SOC is treated here in the context of an effective theory as discussed in Ref. \onlinecite{zhang2020real}. The Hamiltonian then reads as follows:

\bea
\nonumber
\mathcal{H}&=&\mathcal{H}_{t}+\mathcal{H}_{J},\\
\nonumber
\mathcal{H}_t&=&\sum_{\langle j,k\rangle}\left(t_{kj}\mathbf{c}_{k}^{\dagger}U_{kj}\mathbf{c}_{j}+t^{*}_{kj}\mathbf{c}_{j}^{\dagger}U_{kj}^{\dagger}\mathbf{c}_{k}\right),\\
\mathcal{H}_J&=&-\frac{JS}{2}\sum_{j}\mathbf{c}_{j}^{\dagger}\bm{\sigma}\mathbf{c}_{j}\cdot\mathbf{n}_{j}.
\label{eq:hamilto}
\eea
The $\mathbf{c}_{j}$ and $\mathbf{c}_{j}^{\dagger}$ are two-component spinors:
\be
\nonumber
    \mathbf{c}_{j} = \begin{pmatrix}c_{j,\uparrow} \\
                      c_{j,\downarrow}
                      \end{pmatrix} ,  
    \mathbf{c}_{j}^{\dagger} = \begin{pmatrix}
           c_{j,\uparrow}^{\dagger}, c_{j,\downarrow}^{\dagger}
                      \end{pmatrix}.
\ee
The localized spin on site $j$ is represented by the vector $\mathbf{n}_{j}$ which takes three possible values leading to a tripartite lattice as shown in Fig. \ref{fig:red}. The SOC is introduced effectively in the SU(2) matrix $U_{kj}$ through two parameters for each bond connecting the sites $k$ and $j$. Since $U_{kj}$ represents a rotation of the electronic spin we can write it in terms of a unit vector $\mathbf{a}_{kj}$ that gives the direction of the rotation and the rotation angle $\alpha_{kj}$ (that measures the SOC strength):
\be
U_{kj}=\exp\left[-\frac{i\alpha_{kj}}{2}(\mathbf{a}_{kj}\cdot\bm{\sigma})\right],
\ee
where $\bm{\sigma}=(\sigma_{x},\sigma_{y},\sigma_{z})$ is a vector of Pauli matrices. In this effective approach the direction of the vectors $\mathbf{a}_{kj}$, that characterize the SOC, can be inferred following the rules for the Dzyaloshinskii-Moriya interaction (DMI) \cite{moriya1960anisotropic}. According to these rules, the mirror plane $\mathcal{M}_z$ parallel to the lattice plane would set the $\mathbf{a}_{kj}$ to be perpendicular to the lattice plane, but we will consider the lattice to be embeded in a three dimensional structure which breaks this $\mathcal{M}_z$ symmetry. The presence of a mirror symmetry plane $\mathcal{M}_i$ perpendicular to each black bond in Fig. \ref{fig:red} and passing through it's midpoint sets the vectors $\mathbf{a}_{kj}$ to be parallel to $\mathcal{M}_i$ in such bonds. Thus, we choose the vectors $\mathbf{a}_{kj}$ to point to the outer region of the triangular plaquette and $\mathbf{a}_{kj}\perp\mathbf{\hat{r}}_{kj}$, where $\mathbf{\hat{r}}_{kj}$ points in the direction of the intracell bond connecting the sites $k$ and $j$ (black lines in Fig. \ref{fig:red} (a)), that is

\begin{align*}
	 \mathbf{a}_{12} &= \frac{1}{2}\cos(\phi)\hat{x}-\frac{\sqrt{3}}{2}\cos(\phi)\hat{y}+\sin(\phi)\hat{z}, \\
	 \mathbf{a}_{23} &= -\cos(\phi)\hat{x}+\sin(\phi)\hat{z},   \\
	 \mathbf{a}_{13} &= \frac{1}{2}\cos(\phi)\hat{x}+\frac{\sqrt{3}}{2}\cos(\phi)\hat{y}+\sin(\phi)\hat{z}. 
\end{align*}
Where $\phi$, is the angle between the vectors $\mathbf{a}_{ij}$ and the lattice plane. 

Up to now this model corresponds to the general CT lattice and could be the scenario for a wide variety of phenomena if we consider different setups for the parameters. 
However we are going to limit ourselves to a particular case. We consider $|t_1|\leq|t_2|$. The reason for this is twofold. On one side, we found that for $|t_1|>|t_2|$ the Hall conductivity is strongly suppressed even in the presence of a SOC as shown in Fig. \ref{HallCondComparacion}. On the other side, the setup with $|t_1|\leq|t_2|$ has the potential to describe real systems as we discuss in Sec. \ref{sec:realizations}.

While the band structure changes with the value of $\phi$ the topological characteristics of each band remains unchanged so, for concreteness, we set $\phi=0$ for the rest of the article. This situation can reproduce the emergent SOC due to inversion symmetry breaking in real systems (discussed in Sec. \ref{sec:realizations}).

The $120\degree$ magnetic configuration is degenerated for the case of a inversion-symmetric antiferromagnet with pure exchange interactions. This degeneracy corresponds to the two possible chiralities for the $120\degree$ structure (Fig. \ref{fig:red} b), c)). However the presence of an antisymmetric interaction such as the DMI can lift this degeneracy and select a state with definite chirality. For the present case we are going to assume that the magnetic configuration corresponds to that shown in Fig. \ref{fig:red} b). 

%

We describe the properties of this model through the Hamiltonian in Eq. \ref{eq:hamilto} that in momentum space is a $6\times6$ matrix given by:

\be
\mathcal{H}_{\mathbf{k}}= \begin{pmatrix} -\frac{JS}{2} \boldsymbol{\sigma}\cdot \mathbf{n}_{1} & H_{12}^{\dagger} & H_{13} \\ H_{12} & -\frac{JS}{2} \boldsymbol{\sigma}\cdot \mathbf{n}_{2} & H_{23}^{\dagger} \\ H_{13}^{\dagger} & H_{23} & -\frac{JS}{2} \boldsymbol{\sigma}\cdot \mathbf{n}_{3} \end{pmatrix}
\label{eq:bloch_H}
\ee
with 
\begin{align*}
	H_{13} &= t_1U_{13}+t_2\left( e^{-i\mathbf{k}\cdot(\boldsymbol{\delta_{1}}+\boldsymbol{\delta_{2}})}+e^{-i\mathbf{k}\cdot \boldsymbol{\delta_{2}}} \right),\\
	H_{12} &= t_1U_{12}+t_2\left( e^{-i\mathbf{k}\cdot(\boldsymbol{\delta_{1}}+\boldsymbol{\delta_{2}})}+e^{-i\mathbf{k}\cdot \boldsymbol{\delta_{1}}} \right),\\
	H_{23} &= t_1U_{23}+t_2\left( e^{-i\mathbf{k}\cdot\boldsymbol{\delta_{2}}}+e^{i\mathbf{k}\cdot \boldsymbol{\delta_{1}}} \right),
\end{align*}
and $\boldsymbol{\delta}_{1,2}$ the primitive lattice vectors of the Bravais lattice as depicted in Fig. \ref{fig:red}.
%



The most general electronic structure consists of six bands that, for $J=0$ and $\alpha=0$, merge into a lower number of degenerated bands. We will study some particular limits of the previous model and, in order to do that, we are going to consider $t_1+t_2=1$ and parametrize the imbalance between them as $t_1=(1+\lambda)/2$ and $t_2=(1-\lambda)/2$ with $-1\leq\lambda\leq1$.  In the first place we have the case $\lambda=1$ ($t_2=0$). In this situation the system consists of a series of independent triangular plaquettes, this leads to electronic states localized at each plaquette. In fact the band structure corresponds to two flat bands, the upper band is doubly degenerated and the lower band has a fourfold degeneracy. For $0<\lambda<1$ the lower band splits into two bands with two crossing points at the $\Gamma$ and $K$ points. However they remain well separated from the upper band.
A gap closing for $\lambda=0$ ($t_{1}=t_{2}$) introduces additional crossings, now between the upper and lower bands.

On the opposite side $\lambda=-1$ ($t_1=0$), and for $J=0$ and $\alpha=0$, the band structure resembles that of the kagome lattice with three (doubly degenerate) bands and level crossings at the $\Gamma$ and $K$ points. For $-1<\lambda<0$ those band crossings survive and no other crossings are developed insofar $\lambda$ belongs within the previous range.

It is important to mention that since in the model considered here the SOC acts through the $t_1$ terms, in the limit $t_1=0$ the SOC has no effect on the band structure irrespective of the value of $J$. Out of this limit, the introduction of the Hunds coupling and the SOC can induce a band structure in which all bands are well separated from each other. In the next section we are going to study this regime, that is $-1<\lambda<0$, and discuss the effect of a finite SOC on the Hall conductivity.

\section{Hall conductivity}
\label{sec:phase_diagram}
%

Since the Hall conductivity $\sigma_{xy}$ is related to the topology of the electronic bands, we start with a topological analysis of the band structure of the system. To this end we consider the Bloch Hamiltonian $\mathcal{H}_{\mathbf{k}}$ in Eq. \ref{eq:bloch_H}. Without loss of generality, we fix the SOC strength setting $\alpha=0.2\pi$ in the rest of the article. 

As we mention in the previous section, when we move the values of the parameters some of the gaps are closed and then reopened signaling a topological phase transition. 
To characterize this transitions we calculate the Chern numbers $C_n$, where $n$ labels the bands from the lowest ($n=1$) to the highest ($n=6$), associated to each band. The set of Chern numbers can be computed as follows
\be
C_{n}=\frac{1}{2\pi i}\int_{S}\Omega^{n}(\mathbf{k})d^{2}k,
\label{eq:cherns}
\ee
where the integral is evaluated on the surface $S$ corresponding to the Brillouin zone. The Berry curvature is expressed in terms of the Berry connection $A^{n}_{k_{\mu}}(\mathbf{k})$ for the $n$-th band, with $\mu=x$, $y$, through the equation
\bea
\Omega^{n}(\mathbf{k})&=&\partial_{k_{x}}A^{n}_{k_{y}}(\mathbf{k})-\partial_{k_{y}}A^{n}_{k_{x}}(\mathbf{k}),\\
A^{n}_{k_{\mu}}(\mathbf{k})&=&\langle n(\mathbf{k})|\partial_{k_{\mu}}|n(\mathbf{k})\rangle.
\eea
The Chern numbers where computed numerically discretizing the Brillouin zone and computing the Berry Phase in each discrete plaquette in $k$ space, following the numerical method developed by Fukui-Hatsugai-Suzuki\cite{fukui2005chern}.

\begin{figure}[]
\centering
\subcaptionbox{\label{HCP1}}
{\includegraphics[width = \linewidth]{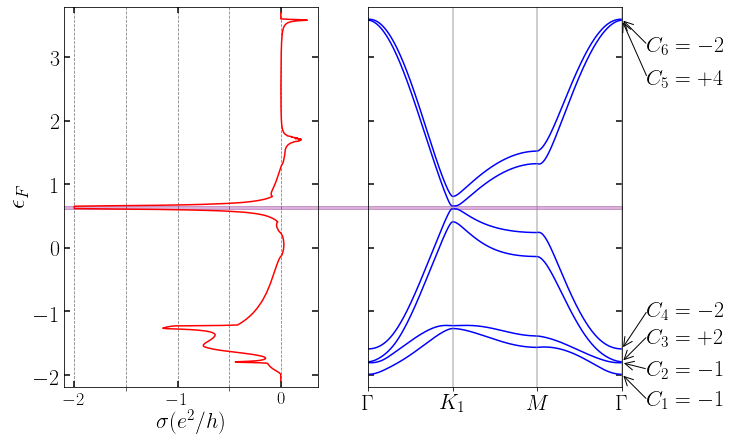}}
\subcaptionbox{\label{HCP4}}
{\includegraphics[width = \linewidth]{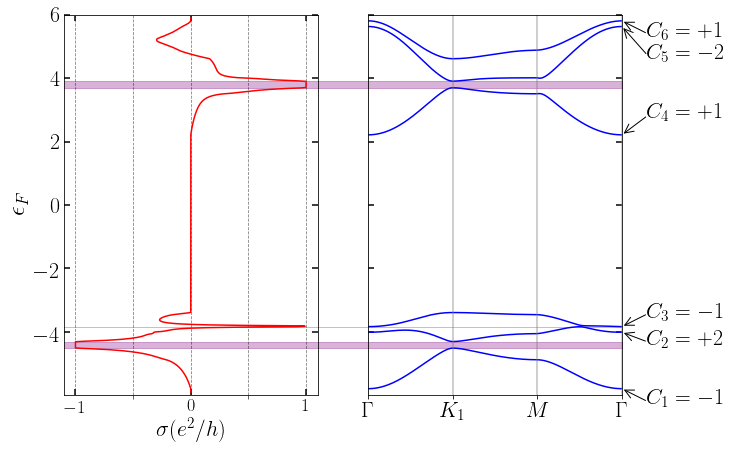}}
\caption{a) Hall conductivity and band structure for $\lambda=-0.6$ and $J=0.4$. One gap is visible in the band structure between bands $4$ and $5$ which leads to a plateau in the Hall conductivity equal to the accumulated Chern number. All bands have a non zero chern number b) Hall conductivity and band structure for $\lambda = -0.6$ and $J=8$. In this case there are three gaps in the band structure, each one leads to its corresponding plateau in the Hall conductivity. All $6$ bands have a non zero chern number. }
\label{HC}
\end{figure}

As mentioned in the previous section, for positive values of $\lambda$ ($t_1>t_2$) the system is topologically trivial, that is we found $C_n = 0$ $\forall n$. For negative values of $\lambda$ we found non-zero Chern numbers for the different bands.  
For example, in Fig. \ref{HC} we show the band structure together with the Chern numbers of each band for two representative cases. 
When the 120$\degree$ structure is uniformly rotated through an angle $\theta$ around a vector perpendicular to the plane of the system, the band structure changes. The Hall conductivity for a Fermi level within a band gap behaves as a step function of $\theta$ (varying from 1 to -1), similar to the behavior shown in Ref. \onlinecite{busch2020microscopic}. Each band retains its non-trivial character regardless of the angle value, although the Chern number sign changes. Thus, without loss of generality, for the magnetic background we keep the structure as depicted in Fig. \ref{fig:red} (b).


So far we have seen that the presence of a SOC could lead to non-trivial bands for suitable ratios of $t_1/t_2$. We analyze now the consequences of those results in the anomalous Hall conductivity, which is our ulterior objective. The Hall conductivity $\sigma_{xy}$ is given by \cite{vanderbilt2018berry}
\begin{equation}
	    \sigma_{xy} = \frac{e^2}{(2\pi)h} \int_{B.Z.} \sum_{n} f(E_{n}(\mathbf{k}))\Omega^{n}(\mathbf{k})d^2k,
\end{equation}
where the sum runs over all energy bands and $f(E_{n}(\mathbf{k}))$ is the Fermi-Dirac distribution. At zero temperature we can approximate $f(E_{n}(\mathbf{k}))$ by a step function $\Theta(\epsilon_f-E_{n}(\mathbf{k}))$, with $\epsilon_f$ the Fermi level, and compute the Hall conductivity as a function of $\epsilon_f$ for different values of $J$ at fixed values $\lambda=-0.6$ and $\alpha=0.2\pi$ as seen in Fig. \ref{HC} for two representative cases. 

\begin{figure*}[htb]
\includegraphics[width=14.5cm]{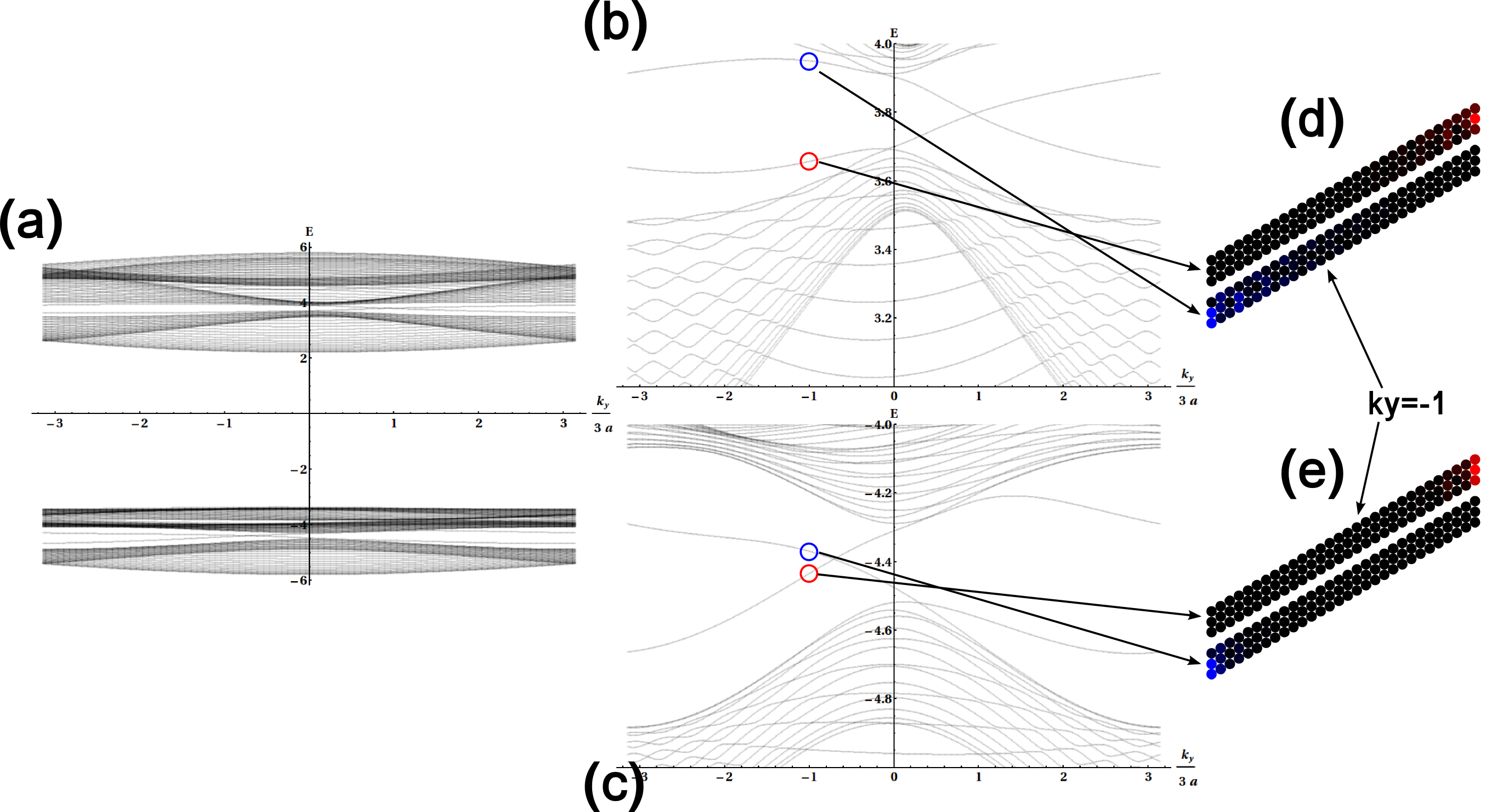}
\caption{(a) The band structure for a ribbon of 30 sites long. (b) Detail of the band structure at the upper band gap (around $E\approx3.8$). (c) Detail of the band structure at the lower band gap (around $E\approx-4.4$). Edge states at $k_{y}=-1$ in the upper (d) and lower (e) band gaps. With red (for right edges) and blue (for left edges) we indicate the square of the wave amplitude at the corresponding site on the ribbon.}
\label{fig:edge_states}
\end{figure*}

We can see that the ingredients considered here could introduce large contributions to the Hall conductivity as a consequence of non-trivial bands. It is interesting to note that even in the weak Hund's coupling regime the Hall conductivity is large. In fact, for small values of $J$ the Hall conductivity could be increased by a factor two as compared with the large $J$ case shown in Fig. \ref{HC} (b). A quantized Hall conductivity is obtained at the global gaps indicated by the red bars in Fig. \ref{HC}, $\sigma_{xy}=C\times e^{2}/h$ where $C=\sum_{n}C_{n}$ and the index $n$ runs over the band index up to the last band below the gap under consideration. This quantization of the Hall conductivity is a hallmark of the non-trivial topological structure of the electronic bands and suggests the presence of robust edge states. The magnitude of the plateaus in the  Hall conductivity is proportional to the number of chiral edge channels. We therefore expect two homochiral states for the case represented in Fig. \ref{HC} (a) at $E\approx0.6$. In the case represented in Fig. \ref{HC} (b) we have at most one chiral state and the chirality of the states at $E\approx-3.9$ and $E\approx4.75$ is opposite to that of the state at $E\approx-4.5$. To show this in this last case we consider a ribbon of 30 sites long (along the direction $\hat{v}=\frac{\sqrt{3}\hat{x}}{2}+\frac{\hat{y}}{2}$) and infinitely long in the $\hat{y}$ direction. The band structure for the ribbon is shown in Fig. \ref{fig:edge_states} (a). The band structure consists in two band bundles separated by a large band gap. There are three smaller gaps, and within these last gaps we can find the aforementioned edge states. In Fig. \ref{fig:edge_states} (b) we show the band structure around the gap at $E\approx4.75$. The bands crossing the gap connects the upper and lower levels, and correspond to gapless edge states. Each of them are localized at opposite ends of the ribbon as shown in Fig. \ref{fig:edge_states} (d) for $k_{y}/3a=-1$. A similar situation is observed for the gap at $E\approx-4.5$ in Fig. \ref{fig:edge_states} (c), where we find again a pair of states crossing the gap and they correspond to edge states as seen in Fig. \ref{fig:edge_states} (e).

\begin{figure}[]
	\includegraphics[width = \linewidth]{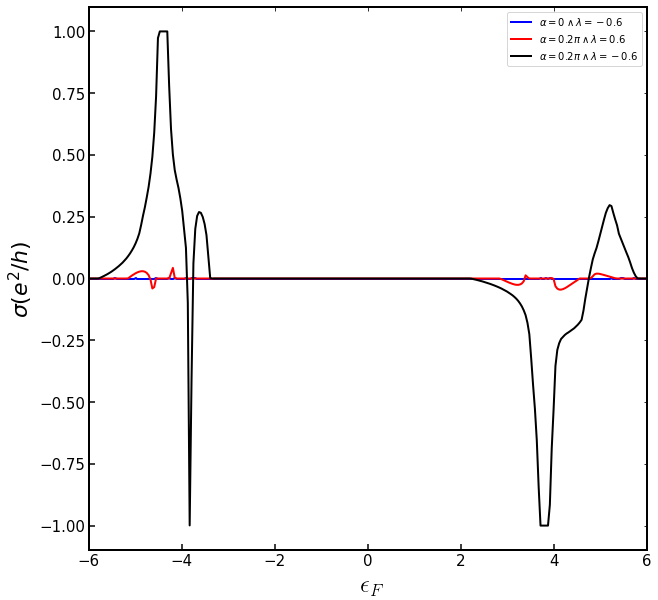}
	\caption{Hall conductivity for three different configurations of the parameters $\lambda$ and $\alpha$ but with a set hund coupling $J=8$. When the spin orbit coupling strength is set to $\alpha=0$ the Hall conductivity is null. If we increase the spin orbit coupling strength the Hall conductivity becomes non-zero but acquires significant values only for $\lambda \leq 0$, which corresponds to $t_2\geq t_1$.}
	\label{HallCondComparacion}
\end{figure}

Since for $\lambda>0$ all the Chern numbers are zero the absolute value of the Hall conductivity becomes orders of magnitude lower than the $\sigma(\epsilon_f)$ obtained for $\lambda<0$, as seen in Fig. \ref{HallCondComparacion}. A different situation arises when we turn off the spin orbit coupling by setting $\alpha = 0$, in this case both $C_n=0$ $\forall n$ and $\sigma_{xy}(\epsilon_f) = 0$ (see Fig. \ref{HallCondComparacion}, blue line), which evidences the relevance of a finite SOC.

%
%
%

%
%

%
\section{Possible realization in real systems}
\label{sec:realizations}
%


Besides the cases discussed previously, we have seen that even for the case of $t_{1}=t_{2}$ the Hall conductivity exhibits a strong plateau. This brings the possibility to realize the anomalous Hall effect in a conventional triangular antiferromagnet, provided the SOC can suitably be introduced on each triangular plaquette. This effect can be induced by the adsorption of transition metal atoms sitting on the lattice sites and voids of a transition metal dichalcogenide monolayer, and forming a triangular arrangement. In Ref. \onlinecite{fang2021spirals} this process actually leads to a DMI between the magnetic moments that form a tringular antiferromagnetic lattice. Thus we expect that the SOC, as considered here, can be induced by this mechanism.
Following the previous line, the more generic CT lattice (with $t_{1}\neq t_{2}$) can be achieved by the same mechanism of adsorption. This was considered in Ref. \onlinecite{zhang2019kagome}. The authors showed that Au atoms adsorbed onto a monolayer of the  compound Ca$_2$N leads to a band structure that has the structure of the kagome bands, and this requires $|t_1|<|t_2|$ in the CT lattice partner.

The magnetization configuration (the 120$\degree$ state) considered here is characteristic of several frustrated antiferromagnets. In particular, the triangular antiferromagnetic lattice exhibits this state as a ground state. However, in the pure Heisenberg Hamiltonian with only exchange interactions, there are many states within this class of configuration. In addition to the SO(3) global symmetry, that corresponds to the different orientations of the planar magnetic configuration, there is an additional degree of freedom, namely, the chirality. This property further distinguishes the 120$\degree$ to belong in two classes as shown in Fig. \ref{fig:red} b) and c). For the pure Heisenberg Hamiltonian the two chiralities are degenerated. However, the chirality selection can occur due to chiral interactions such as the DMI. This kind of process could take place, for example, in the triangular antiferromagnet  Ba$_3$NbFe$_3$Si$_2$O$_{14}$  \cite{zorko2011role}. It is important to mention that in a recent article the authors claim that the DMI is not strong enough to lift the chirality degeneracy \cite{qureshi2020absolute}.

Thus, the different and key ingredients of our model are present in actual electronic and magnetic systems. So far, we have considered them as separated components and spread over a wide range of materials. Reuniting them into a single compound can be challenging, since these ingredients can appear entangled in a real system, which possibly can make it difficult to obtain a representative material consistent with our model. However, as we mentioned earlier, the emergent band structure of the CT lattice resembles that of the kagome lattice (in fact by setting $t_{1}=0$ the band structure of the kagome lattice is recovered), which provides an alternative scenario in which the previous ingredients could be disentangled and effectively realize the CT lattice as considered here \cite{zhang2019kagome}.

\vspace{0.7cm}

\section{Conclusions}
\label{sec:conclusions}

We have studied the anomalous Hall effect on a CT lattice antiferromagnet and showed that a strong anomalous Hall effect is present in a coplanar magnetic texture without net magnetization. The SOC is at the heart of this phenomena since, for the system without SOC, the Hall conductivity is inhibited. Another key ingredient is the different values for the intra-plaquette ($t_1$) and the inter-plaquette hoppings ($t_2$). When the former is weaker than the latter, a strong Hall conductivity emerges. The other way around, for $t_1>t_2$, the Hall conductivity is strongly suppressed. 

The results discussed here reveal that the CT lattice provides an excellent playground for the anomalous Hall effect even in the antiferromagnetic case with a coplanar magnetic configuration. As such, we expect that the phenomena described here could be realized in the systems previously discussed, from which we highlight the antiferromagnetic organic compounds with the structure of the CT lattice and triangular antiferromagnetic systems formed with transition metal atoms adsorbed onto a transition metal dichalcogenide monolayer. In view of the similarity between the band structure of the kagome lattice and the CT lattice, the latter also provides an excellent playground for studying the interplay between strong correlations and topology in electronic systems.

\appendix
%
%
\section*{Aknowledgments}
We would like to thank Cristian Batista for useful discussions and suggestions.

\bibliography{references.bib}

\end{document}